# Does Monetary Support Increase Citation Impact of Scholarly Papers?[1]

Yaşar Tonta[1] and Müge Akbulut[2]

[1]yasartonta@gmail.com
Hacettepe University, Faculty of Letters, Department of Information Management, Ankara (Turkey)

[2]mugeakbulut@gmail.com
Ankara Yıldırım Beyazıt University, Faculty of Humanities and Social Sciences, Department of Information Management, Ankara (Turkey)

**Abstract**
One of the main indicators of scientific development of a given country is the number of papers published in high impact scholarly journals. Many countries introduced performance-based research funding systems (PRFSs) to create a more competitive environment where prolific researchers get rewarded with subsidies to increase both the quantity and quality of papers. Yet, subsidies do not always function as a leverage to improve the citation impact of scholarly papers. This paper investigates the effect of the publication support system of Turkey (TR) on the citation impact of papers authored by Turkish researchers. Based on a stratified probabilistic sample of 4,521 TR-addressed papers, it compares the number of citations to determine if supported papers were cited more often than those of not supported ones, and if they were published in journals with relatively higher citation impact in terms of journal impact factors, article influence scores and quartiles. Both supported and not supported papers received comparable number of citations per paper, and were published in journals with similar citation impact values. Findings suggest that subsidies do not seem to be an effective incentive to improve the quality of scholarly papers. Such support programs should therefore be reconsidered.

**Introduction**
The number of refereed papers that appears in scientific journals along with citations thereto is considered to be the main indicators of scientific productivity and quality of a given researcher, a research organization or a country. Many countries introduced what is called performance-based research funding systems (PRFSs) to streamline the scientific production process and improve the research performance (Jonkers & Zacharewicz, 2016).

PRFSs aim to assess the performances of researchers in a given time period. Some countries provide monetary incentives directly to researchers in the form of "piece rates" or "cash-for-publication" schemes (Heywood, Wei, & Ye, 2011) while others prefer to reward researchers' organizations by allocating funds to them (De Boer et al., 2015). Both "ex ante" and "ex post" assessments are being used for this purpose. Compared with peer review requiring labor-intensive evaluation processes prior to funding allocation, it is relatively easier, and less costly, to carry out ex post quantitative assessments on the basis of bibliometric measures.

Notwithstanding the type of assessment carried out, research organizations or countries tend to eagerly incentivize their researchers because they in turn expect return on investment (RoI), usually as an increase in the number of papers published by their researchers as well as the citation impact of their papers. However, such monetary incentives do not necessarily produce the intended outcomes, as the existence of PRFSs does not correlate well with the research productivity or quality (Auranen & Nieminen, 2010). The effect of PRFSs on the increase in the quantity of publications is "temporary and fades away after a few years" while the average effect on the quality of publications is "nil" (Checchi, Malgarini, & Sarlo, 2019, pp. 45, 59).

---

[1] This is a revised and expanded version of a paper presented at ISSI 2019: 17th International Scientometrics and Informetrics Conference, 2-5 September 2019, Sapienza University of Rome, Italy (Tonta & Akbulut, 2019).

This paper aims to study the effect of the publication support system of the Turkish Scientific and Technological Research Council (TÜBİTAK) on the increase of the citation impact of papers published in scientific journals. The support system is based on the concept of "cash-for-publication" and has been in place since 1993. The authors of papers get rewarded on the basis of Journal Impact Factors (JIFs) and, more recently, Article Influence Scores (AISs) of journals in which their papers are published. The higher the JIF or AIS values, the more money the authors get paid, with a cap of 7,500 Turkish Lira (circa 3,000USD in 2015), although papers with no JIFs or AISs get supported as well (TÜBİTAK, 2015, p. 4), presumably to increase the number of TR-addressed papers in Social Sciences and especially in Arts and Humanities. We compare all papers published between 2006 and 2015 and indexed in Web of Science (WoS) with at least one author whose address is based in Turkey (henceforth "TR-addressed papers") with those supported by TÜBİTAK to see if supported papers received more citations and if they were published in higher quality journals in terms of JIFs, AISs and quartiles (Q1 through Q4).

The rest of the paper is organized as follows: The next section briefly reviews relevant studies. The Data Sources and Method section describes the data and sampling technique used to select the TR-addressed papers along with the matching algorithms written to identify the supported ones. The Findings and Discussion section presents detailed findings. The paper ends with Concluding Remarks.

**Literature Review**

Performance-based research funding systems (PRFS) based on monetary rewards have been introduced in 1980s (Quan, Chen, & Shu, 2017). The rationale behind PRFSs is to reward researchers or institutions with higher performances (in terms of publishing more papers, for instance) the ones with lower performances strive to improve themselves to get rewarded (Herbst, 2007). The Research Excellence Framework (REF) of UK is one of the oldest PRFSs based on peer review and has been in use since 1986 (De Boer et al., 2015, p. 113). Yet, bibliometric measures such as JIF have become the dominant method used for research evaluation purposes within the last two decades and they are readily available through Journal Citation Reports (JCR) published annually by Clarivate Analytics. JIF coefficients are sometimes applied to calculate the amount of payment formulaically (Quan, Shen, Chu, 2017; Shao & Shen, 2012; Sombatsompop & Markpin, 2005, pp. 676-677). JIF seems to be so popular in some countries (e.g., China) that authors who publish in indexed journals get up to 30% discount in local restaurants (the greater the JIF, the higher the discount) (Hongyang, 2017). However, there is no correlation between the citation impact of a paper and its JIF (Zhang, Rousseau, & Sivertsen, 2017), and the disadvantages of using JIF as a performance indicator have been reviewed in a number of studies (e.g., Casadevall & Fang, 2012; Glänzel & Moed, 2002; Marx & Bornmann, 2013; Seglen, 1997; Wilsdon et al., 2015; Wouters et al., 2015).

Nonetheless, several countries have developed primarily JIF-based PRFSs where JIF values are used (sometimes in combination with peer review) to determine the amount of monetary support per paper. The PRFS use around the world has been reviewed in a number of studies (e.g., Auranen & Nieminen, 2010, De Boer et al., 2015; European Commission, 2010; Geuna & Martin, 2003; Hicks, 2012; Pajić, 2014). The practices tend to vary from country to country. Some reward researchers directly through what is called "cash-for-publication" schemes (e.g., China and Turkey) while others support the affiliated research units or universities (e.g., UK and South Africa) (Hedding, 2019; Heywood, Wei, & Ye, 2011; Lee & Simon, 2018; Quan, Chen, & Shu, 2017; Tonta, 2017a).

PRFSs tend to produce unintended consequences or "side effects" (Geuna & Martin, 2003) and cause researchers to develop "opportunistic behaviors" (Abramo, D'Angelo & Di Costa, 2019; Good, Vermeulen, Tiefenthaler, & Arnold, 2015). For instance, due to the cash reward policies instituted in early 1990s in China, researchers were faced with "the golden rule of academia in China: publish or impoverish", and they were eager to produce quick and "cashable" papers. As the stakes are quite high,[2] some researchers resort to "plagiarized or fabricated research, purchase ghostwritten papers, or sell authorship" (Quan, Chen, & Shu, 2017, pp. 498-499). Hence, PRFSs have eventually become "perverse incentives" (Tomaselli, 2018).

Some researchers tend to target metrics such as number of citations and h-index measuring the quantitative impact of their research and manipulate them to their advantage, thereby making such metrics useless and further reinforcing the Goodhart's Law ("When a measure becomes a target, it ceases to be a good measure"). The results of a recent large-scale analysis of more than 120 million scientific papers seem to support Goodhart's Law (Fire & Guestrin, 2019). Moreover, counting seems to change what is counted and metrics tend to "commodify and disempower" what they measure (Sætnan, Tøndel, & Rasmussen, 2019, p. 81).

Muller (2017) studied the subsidies from the viewpoint of "rent seeking" theory in economics and explored the impact of distorted incentives on academia, academics and society at large. According to rent seeking theory, academics "compete for artificially contrived transfers" in various forms (e.g., grant funding, monetary incentives for publications and citations). These transfers are usually redirected by public institutions from social surplus to rent seeking academics on the basis of bibliometric measures that are thought to measure academic success better and "provide greater reassurance of quality" (p. 59). Such measures are therefore increasingly supplanting (rather than supporting) peer review used to judge the quality of scholarly output, and universities are "creating institutional rules and practices that actively incentivize rent-seeking behavior" (Muller, 2017, p. 61). In South Africa, for instance, the amount of monetary support per paper (which may be as high as 10,000USD per a single-authored paper) is the same, regardless of where the paper has been published as long as the outlet is "accredited" by the Department of Higher Education and Training. Thus, one can submit their work to lower quality journals with relatively lower standards of peer review in order to collect subsidies quickly and more often. This may, in turn, have created a "powerful perverse incentive" and encouraged at least some researchers to "game" with the system and produce "fraudulent –or ethically questionable– publications" (Muller, 2017, pp. 63-64). Perverse effects of "citation gaming" can even be detected in country-level citation analysis studies (Baccini, De Nicolao, & Petrovich, 2019). Muller (2017) underlines the dilemma of such incentives as follows:

> Under the rent seeking conceptualization of such systems, appeals to individual or institutional integrity are not likely to be successful. The system directly creates incentives for the activities cautioned against, undermining cultures of ethical practice, and therefore only measures that carry suitable material punishment are likely to counteract these undesirable effects. (p. 64)

The side effects of PRFSs are not limited only with researchers publishing in lower quality outlets or "seeking out 'easier' publication types" (Sīle & Vanderstraeten, 2019, p. 86). Subsidies tend to discourage types of research that require more time to carry out using novel experiments prolonging the publication process, thereby giving way to papers with little or no societal impact whatsoever (Geuna & Martin, 2003; Tonta, 2017b, pp. 27-30). Moreover, some researchers simply prefer to publish in "predatory journals" and set up what is called "citation circles" to benefit more from PRFS (Good, et al., 2015; Teodorescu & Andrei, 2014, pp. 228-229). As payouts tend to encourage professors to publish in predatory journals, South African researchers, for instance, published as much as five times more papers in them than those in the United States or Brazil did (Hedding, 2019). While the number of South African publications has doubled (as pointed out earlier) after the introduction of the subsidy program, the ones published in predatory journals increased 140 times during the same period (2004-2010) (Mouton & Valentine, 2017).

---

[2] A professor may earn the equivalent of 20 years' salary for a *Nature* or *Science* paper, and the maximum amount can be as high as 165,000USD for a single paper (Quan, Chen, & Shu, 2017, pp. 491, 494).

Turkey does not have a very good reputation, either, as it ranks third (after India and Nigeria) in the world among 146 countries in terms of number of papers published in predatory journals (Demir, 2018a, p. 1303). Beall's (now defunct) list of predatory journals includes 41 such journals originating from Turkey, second highest after India (Akça & Akbulut, 2018, p. 264). Some researchers indicated that the academic incentive system of the Turkish Higher Education Council (HEC) that has been in place since 2016 is one of the factors encouraging them to publish in predatory journals (Demir, 2018a, p. 1307). Not surprisingly, the number of TR-addressed papers published in predatory journals and presented in dubious conferences has significantly increased since then (Demir, 2018b, pp. 2057-2058). HEC has recently taken some measures to curtail such attempts by researchers but they are not stringent enough. Fortunately, papers published in predatory journals would no longer count towards tenure and promotion (Koçak, 2019, p. 200). However, it is not yet totally clear at this point in time as to which criteria to apply in order to identify such papers and exclude them from tenure and promotion process, from HEC's academic incentive system as well as from TÜBİTAK's support program. Although no such study has so far been carried out, it seems that the numbers of TR-addressed papers published in predatory journals have already been subsidized in the past by TÜBİTAK under its support program. For example, more than 80% of the subsidies for papers in anthropology in 2015 went to a single predatory journal in this field in which Turkish researchers have published a total of 127 papers, most of which had had nothing to do with anthropology (Tonta, 2017b, p. 80).

Despite the side effects and undesired outcomes of PRFSs, there appears to be a commonly held belief in research funding and research performing institutions that subsidies would increase the number of papers and their citation impact. Researchers motivated by such subsidies would produce more papers with higher quality. However, the relationship between subsidies and the increase in productivity and quality is not clear-cut (Auranen & Nieminen, 2010). While there appears to be some evidence that subsidies increase the number of papers to some extent, this is not reciprocated with a similar increase in the quality of papers in terms of their citation impact (Butler, 2003, 2004; Good et al., 2015; Osuna, Cruz-Castro, & Sanz-Menéndez, 2011). For instance, the number of South African publications has almost doubled in seven years (2004-2010) after the implementation of the subsidy system. Yet, their citation impact (number of citations per paper) has decreased steadily (Pillay, 2013, p. 2). A small-scale study carried out at the University of Cape Town after the implementation of PRFS showed that the number of output is negatively correlated with both the number of citation counts of papers and their field-weighted citation impact. Although the variance explained was relatively modest, findings indicate to some extent that greater subsidy seems to be "associated with lower citation impact," which may, in part, be due to the fact that the PRFS currently in use "does not factor in research quality and impact" (Harley, Huysamen, Hlungwani & Douglas, 2016). Similarly, the number of TR-addressed papers listed in citation indexes has increased 19-fold between 1993 (when TÜBİTAK's support program began) and 2015 (from 1,500 papers to more than 28,000) (Tonta, 2017b, p. 32). Turkey has jumped from 37$^{th}$ place in 1993 to 18$^{th}$ in 2008 in the world ranked by the number of indexed publications. Yet, findings of an interrupted time series analysis based on 390,000 TR-addressed publications listed in the WoS database between 1976 and 2015 (of which 157,000 or 40% were subsidized between 1997 and 2015) showed that the support program seems to have had no impact on the increase in the quantity of TR-addressed publications (Tonta, 2017a). Moreover, the citation impact of TR-addressed papers has constantly decreased throughout the years and is well below (40%) that of the world average of all papers (Çetinsaya, 2014, p. 127; Kamalski et al., 2017, p. 4).

It appears that PRFSs do not help much in terms of improving the quality of research. In fact, PRFSs based on cash rewards are considered to be "the enemy of research quality", as there appears to be a negative correlation between subsidies and the number of citations and field-weighted impact of publications (Hedding, 2019; Harley et al., 2016). The pressure of "publish or perish" tends to force researchers to choose between "cash or quality" (Hedding, 2019), which "is making science perish" along the way as well (Şengör, 2014, p. 44).

We test this conjecture of "cash or quality" with reference to Turkey and see if the subsidy system currently in use in Turkey has improved the citation impact of TR-addressed papers by comparing the number of citations per paper, JIFs, AISs and quartiles of journals in which both supported and not supported papers were published. We present below the data sources and method used to analyze the data.

**Data Sources and Method**

We used the well-known bibliometric measures of number of citations per paper, JIF, AIS and JCR quartiles to compare the citation impact of supported and not supported TR-addressed papers. JIF is defined as the "average" citation impact of papers published in a given journal within a given time period. AIS takes into account five years' worth of citation data for a given journal and weights citations on the basis of JIFs. If citations come from high impact journals, they are weighted more heavily. AIS is similar to Google's PageRank algorithm in that it uses the whole JCR citation network to calculate the AIS for a given journal. Unlike JIS, AIS indicates if each article in a journal has above- or below-average influence, 1.000 being the average of all journals included in JCR's citation network (Article, 2019). AIS is more stable and can therefore be used in interdisciplinary comparisons where journals have varying publication and citation patterns, although both metrics are highly correlated ($r = .9$) (Arendt, 2010).

JIF is used to categorize journals under at least one subject category, and journals under each subject category are divided into four quartiles based on their JIF values as a field normalized indicator (the first 25% of the journals with highest JIFs constitutes Q1, the second 25% Q2, etc.). Some journals are listed under two or more subject categories, hence sometimes under different quartiles. In this case, we categorized such journals under their highest quartiles. Journals in the Q1 quartile are deemed "high impact journals" with impact factors in the top 25% of the JIF distribution of journals in a certain field while the ones in the Q4 quartile representing the bottom 25% are considered "low impact journals". Journal titles are, by definition, more or less evenly distributed to JCR quartiles. Yet, publications are not evenly distributed by JCR quartiles (Liu, Hu, & Gu, 2016; Liu, Hu, & Gu, 2018; Miranda & Garcia-Carpintero, 2019). For instance, according to the 2015 JCR-Science Edition based on more than 8,500 journals, somewhere between 36%-46% of more than 1.3 million publications appeared in high impact Q1 journals compared to that of only 13%-17% that were published in low impact Q4 journals in 2014, and the distributions in other years were similar (Liu, Hu, & Gu, 2016).

We used "articles" as publication type in this study, as some 98% of all TR-addressed publications indexed in Web of Science (WoS) have been of this type. We excluded "reviews" because (1) they made up less than 2% of all TR-addressed publications; (2) the amount of cash paid by TÜBİTAK to the authors of reviews (as well as to those of "notes" and "discussion papers") was half that of articles; and (3) except the number of citations, we use journal-level parameters of JIF, AIS, and JCR quartiles for comparison of supported and not supported TR-addressed papers. As for the number of citations, the citation impact of the exclusion of citations generated by a small number of reviews seems to be negligible.

In order to identify all TR-addressed papers (articles only) published between 2006 and 2015 and indexed in WoS, we used the following advanced search query (December 2, 2017):

**AD**=(Turkey OR Turquie OR Türkei OR Türkiye OR Turquia)
**Timespan**: 2006-2015. **Indexes**: SCI-EXPANDED, SSCI, A&HCI. **PubType**: Article

We found a total of 225,923 TR-addressed papers[3] and downloaded them. We obtained the payment information for 100,919 TR-addressed papers whose authors sought financial support from TÜBİTAK through its Support Program of International Publications (UBYT). Altogether some 44% of all papers were supported (range: 59% in 2007; 28% in 2015).

We then stratified all TR-addressed papers by year and scrambled them within each year (in case they had an inherent order by author or journal name, for example) so that certain records would not appear disproportionally in the sample. We wrote a macro to select every 12th and 75th records (these numbers were randomly chosen) out of the stratified list of all TR-addressed records (225,923).[4] Sample size being 2% of the population, we obtained a total of 4,521 records in the sample using the stratified probability sampling technique. The sample size for each year ranged between 1.86% and 2.05%, average being 1.99% (which is quite close to 2%).

Next, we wrote a second macro to match up journal data from JCR and InCites with respective years to identify bibliometric characteristics of journals (e.g., JIF, AIS, Times Cited, Quartiles and so on) in which TR-addressed papers appeared along with the number of citations that each paper received, if any (February 2, 2018). Data were then added to all the records (seven journals did not match due to inconsistencies in journal names, which were added manually).

Finally, we wrote a third macro to match the list of papers supported by TÜBİTAK with all papers (supported or not). (Some 64 records did not match due to inconsistencies in paper titles, which were added manually.) This enabled us to compare both paper and journal characteristics for both supported and not supported papers (e.g., number of citations for papers, and JIF, AIS and quartile for journals). The matching algorithm seems to have worked quite well. Altogether, 44% of all papers were supported by TÜBİTAK. In the sample, the percentage of supported papers was somewhat lower: 1,679 out of 4,521 (or 37%). The difference (7%) is due to inconsistencies in data (such as punctuation marks and abbreviations used in titles of papers and journals). Nevertheless, we looked at the data more closely. Papers appeared in 9,463 different journal titles. The ones supported by TÜBİTAK appeared in 2,336 different journal titles. Some 2,153 journals (or 92%) were represented in the sample, of which 986 (or 42%) had published at least one supported TR-addressed paper between 2006 and 2015. We do not expect such small fluctuations to have any considerable impact on the analysis that follows.

We used MS Excel and SPSS 23 for data analysis and visualization; independent samples *t*-test for significance, as our sample size was relatively large (4,521) (Lumley, Diehr, Emerson, & Chen, 2002); and chi-square for test of independence. We used an alpha level of .05 for all statistical tests.

**Findings and Discussion**

As indicated earlier, the number of papers published between 2006 and 2015 is 225,923. Table 1 and Fig. 1 provide average and median JIFs and AISs of journals in which all TR-addressed papers appeared during this period. The median JIFs range between 0.998 (2012) and 1.379 (2015) while median AISs range between 0.321 (2012) and 0.457 (2010). Close to half the papers were published in low impact journals (i.e., JIF below 1.000), and their AIS values were less than half (around or below 0.400) of that of the world average (1.000). Although there seems to be a slight increase in recent years in median JIFs and AISs of papers, this has probably more to do with the continuing increase in JIFs over the years (Fischer & Steiger, 2018).

---

[3] In this study, we use "papers" or "TR-addressed papers" in general instead of "articles or "TR-addressed articles", unless otherwise indicated.

[4] We actually selected three different samples (every 50th and 99th record; every 12th and 77th record; and every 12th and 75th record) with the same sample size and compared the descriptive statistics such as means and medians to make sure the stratified probabilistic sampling technique worked properly. As sample statistics were quite similar in all three cases, we report here the findings based on the last one.

Table 1. Average and median JIFs and AISs of journals in which all TR-addressed papers published (2006-2015)

| Year | Average JIF | Average AIS | Median JIF | Median AIS |
|------|-------------|-------------|------------|------------|
| 2006 | 1.481 | 0.508 | 1.087 | 0.411 |
| 2007 | 1.311 | 0.454 | 1.091 | 0.406 |
| 2008 | 1.444 | 0.536 | 1.098 | 0.403 |
| 2009 | 1.409 | 0.481 | 1.072 | 0.399 |
| 2010 | 1.546 | 0.526 | 1.245 | 0.457 |
| 2011 | 1.327 | 0.459 | 1.053 | 0.384 |
| 2012 | 1.401 | 0.455 | 0.998 | 0.321 |
| 2013 | 1.626 | 0.504 | 1.231 | 0.351 |
| 2014 | 1.769 | 0.537 | 1.234 | 0.342 |
| 2015 | 1.988 | 0.574 | 1.379 | 0.368 |

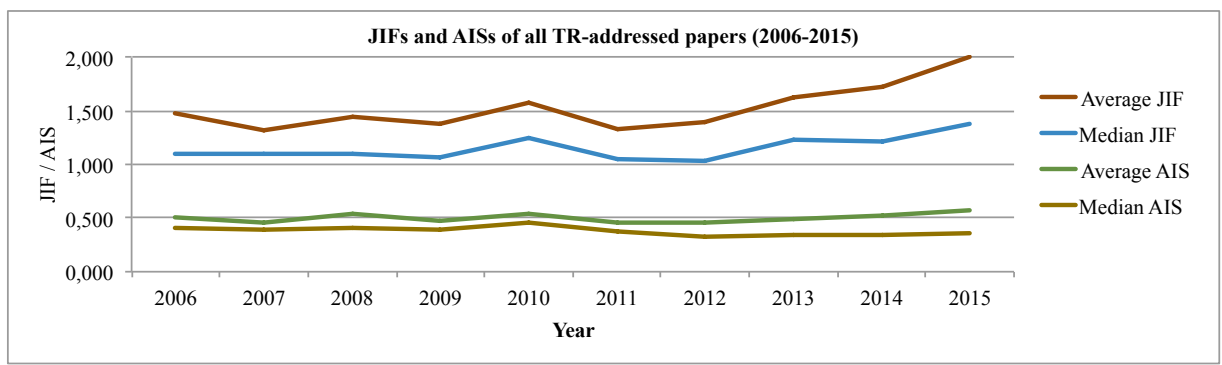

Fig. 1. Average and median JIFs and AISs of journals in which all TR-addressed papers published (2006-2015)

Findings below are based on the stratified probability sample of 4,521 papers (2006-2015). The great majority (90%) of the papers in the sample were Science papers. Social Science and Arts and Humanities papers constituted about 9% and 1% of all papers, respectively. Papers in the sample were cited a total of 55,383 times. The average number of citations per paper was 12 ($SD$ = 42). Half the papers received five or fewer citations (min. = 0, max. = 2,246). Only 1% (or 45 papers) received 100 or more citations while 32% received 10 or more. Some 13% of papers were not cited at all. As expected, Science papers received the overwhelming majority of the total number of citations (over 92%) followed by Social Science papers (7%) and Arts and Humanities papers (less than 1%).[5]

Table 2 provides descriptive statistics and statistical test results for papers in the sample. As indicated earlier, 37% (1,679) of papers in the sample were supported by TÜBİTAK. Supported papers collected 43% (23,654) of all citations. On average, supported papers were cited slightly more often ($M_{all}$ = 14.1, $SD$ = 22.5) than not supported ones ($M_{all}$ = 11.0, $SD$ = 49.8). This difference was significant $t$ (4,519) = -2.39, $p$ < .05; however, the effect size was rather small ($r$

---

[5] Note that 49 Arts and Humanities papers that received a total of 289 citations were excluded from further analysis as bibliometric characteristics of Arts and Humanities journals are not listed in JCR.

= .04). Similarly, supported science papers were cited somewhat more frequently on average ($M_{sci}$ = 14.5, $SD$ = 22.7) than not supported ones ($M_{sci}$ = 11.3, $SD$ = 52.1). Again, although the difference was significant $t$ (4,081) = -2.28, $p$ < .05, the effect size was infinitesimal ($r$ = .04).

Table 2. Descriptive statistics and test results

| | Citation impact | Supported papers | | | | Not supported papers | | | | t | p | r |
|---|---|---|---|---|---|---|---|---|---|---|---|---|
| | | N | Mdn | M | SD | N | Mdn | M | SD | | | |
| All papers | # of cit. | 1679 | 7.0 | 14.1 | 22.5 | 2842 | 4.0 | 11.0 | 49.8 | -2.39 | **.02*** | .04 |
| | JIF | 1624 | 1.2 | 1.6 | 2.3 | 2696 | 1.1 | 1.5 | 2.0 | -1.07 | .28 | .02 |
| | AIS | 1520 | .4 | .5 | .9 | 2405 | .4 | .5 | 0.9 | -.39 | .70 | .01 |
| Science | # of cit. | 1508 | 7.5 | 14.5 | 22.7 | 2575 | 4.0 | 11.3 | 52.1 | -2.28 | **.02*** | .04 |
| | JIF | 1482 | 1.3 | 1.6 | 2.4 | 2473 | 1.1 | 1.5 | 2.0 | -1.17 | .24 | .02 |
| | AIS | 1388 | .4 | .5 | .9 | 2218 | .4 | .5 | .9 | -.48 | .63 | .01 |
| Social Science | # of cit. | 171 | 4.0 | 10.6 | 20.1 | 267 | 3.0 | 8.7 | 13.8 | -1.18 | .24 | .05 |
| | JIF | 142 | 1.2 | 1.3 | 1.0 | 223 | 1.0 | 1.4 | 1.3 | -.43 | .67 | .02 |
| | AIS | 132 | .4 | .4 | .3 | 187 | .4 | .5 | .50 | .52 | .61 | .03 |

*Notes*: *N*: Number of papers; *# of cit.*: Number of citations; *Mdn*: Median; *M*: Mean; *SD*: Standard Deviation; *t*: t-test; *p*: p value; *r*: effect size; *cit.*: citation; *JIF*: Journal Impact Factor; *AIS*: Article Influence Score; *: statistically significant at alpha level .05.

That during a 10-year period a supported paper in general and a supported Science paper in particular received on average about three more citations than a not supported one did, and that the difference was statistically significant, has probably more to do with the sample size than a true effect. This is because insubstantial differences can still be significant with smaller *p* values in a relatively large sample size (as is the case in this study) (Altman & Krzywinski, 2015, p. 900). To put the difference into a better perspective, assuming that any given paper had on average six years to collect citations between 2006 and early 2018, a supported paper received about half a citation more per year than a not supported one did. This can hardly be considered a substantial difference. In fact, the statistically significant difference per paper between the numbers of citations for supported ($M_{ssci}$ = 10.6, $SD$ = 20.1) and not supported ($M_{ssci}$ = 8.7, $SD$ = 13.8) Social Science papers disappeared ($t$ (436) = -1.18, $p$ > .05), as Social Science papers constituted less than one tenth of all papers in the sample. Fig. 2 provides a comparative view of the number of citations per paper for all papers as well as for Science and Social Science papers. Boxplots show the means, medians, and first and third quartile values for both supported and not supported papers.

Table 2 also provides data on citation impact values (e.g., JIF and AIS) of journals in which TR-addressed papers were published. On average, the JIF values of journals publishing all supported papers ($M_{all}$ = 1.6, $SD$ = 2.3), supported Science papers ($M_{sci}$ = 1.6, $SD$ = 2.4), and supported Social Science papers ($M_{ssci}$ = 1.3, $SD$ = 1.0) were quite similar to those publishing not supported ones ($M_{all}$ = 1.5, $SD$ = 2.0; $M_{sci}$ = 1.5, $SD$ = 2.0; and $M_{ssci}$ = 1.4, $SD$ = 1.3, respectively). (The median JIFs of supported and not supported papers were also very close to each other.) The differences were not statistically significant in all three cases ($t_{all}$ (4,318) = -1.07, $p$ > .05; $t_{sci}$ (3,953) = -1.17, $p$ > .05; and $t_{ssci}$ (363) = 0.41, $p$ > .05, respectively).

As for the average AIS values of journals in which TR-addressed supported and not supported papers, they were almost the same and their distributions (i.e., standard deviations) were identical. On average, the AIS values of journals publishing all supported papers ($M_{all}$ = .5, $SD$ = .9), supported Science papers ($M_{sci}$ =.5, $SD$ =.9), and supported Social Science papers ($M_{ssci}$ =.4, $SD$ =.3) were also quite similar to those publishing not supported ones ($M_{all}$ = .5, $SD$ = .9; $M_{sci}$ = .5, $SD$ = .9; and $M_{ssci}$ = .4, $SD$ = .5, respectively). The median AIS values of all, Science, and Social Science papers were the same (.4) regardless of their being supported or not. Not surprisingly, there were no statistically significant differences in all three cases ($t_{all}$ (3,923) = -.39, $p > .05$; $t_{sci}$ (3,604) = -.48, $p > .05$; and $t_{ssci}$ (317) = -.52, $p > .05$, respectively) (Table 2).

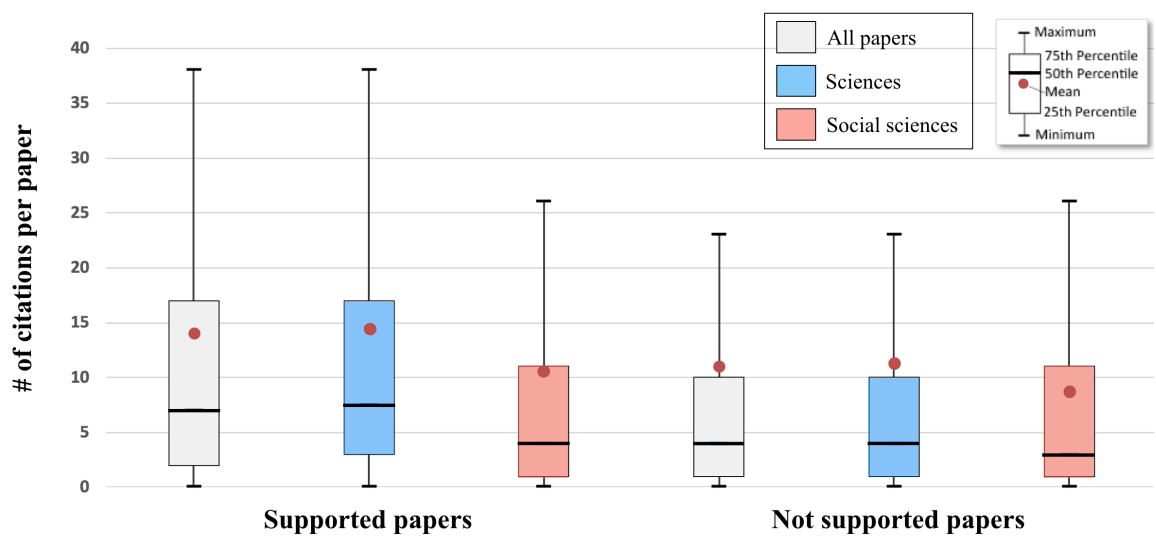

Fig. 2. Number of citations per paper for supported and not supported TR-addressed papers

Fig. 3 below provides the boxplots for JIF and AIS values of journals publishing both supported and not supported papers for all papers as well for Science and Social Science papers. JIF and AIS data of both supported and not supported papers were highly skewed with heavy tails, indicating that papers were mostly published in relatively mediocre or low impact journals. Average JIF and AIS values of journals obtained from the sample are consistent with those of all journals in the population (see Table 1, Fig. 1).

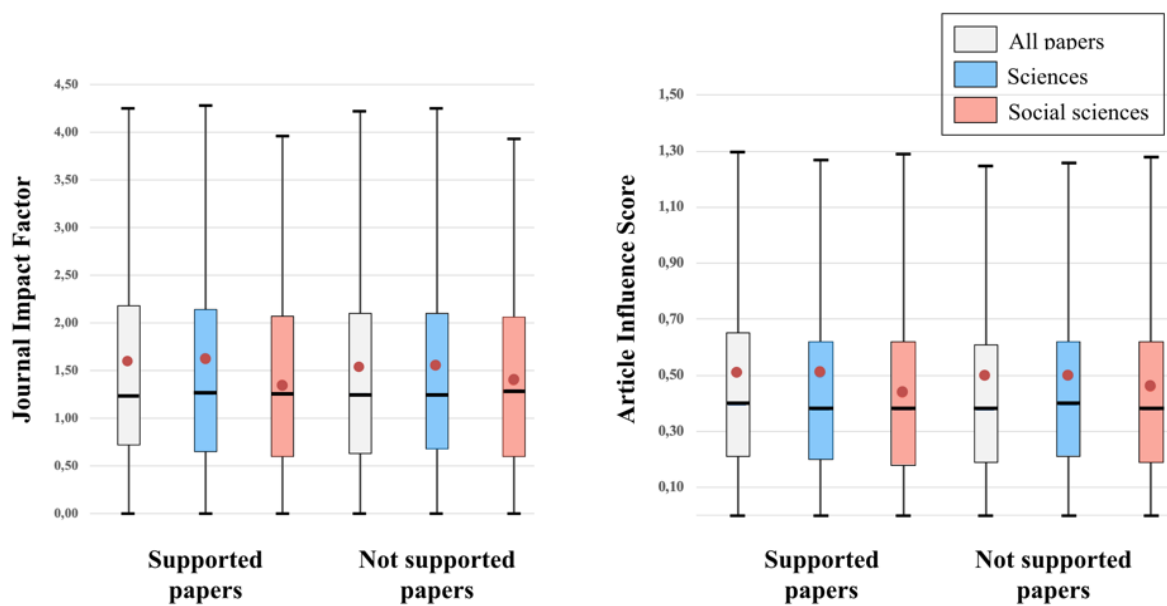

Fig. 3. Journal Impact Factors (left panel) and Articles Influence Scores (right panel) of journals publishing supported and not supported TR-addressed papers

Fig. 4 below provides the percentage distributions of JIF values of supported and not supported papers. Note that the percentages of supported and not supported papers were quite similar to each other, supporting the results of the statistical tests further. Correlation between JIF and AIS values of journals publishing TR-addressed papers was quite high (Pearson's *r* = .946), explaining 90% of the variance in the data[6] and confirming the findings of similar studies (e.g., Arendt, 2010). We therefore do not provide the distributions of JIF and AIS values of supported and not supported Science and Social Science papers separately, as they are similar to those in Fig. 4. Percentages of supported Science and Social Science papers in the sample were 37% and 39%, respectively.

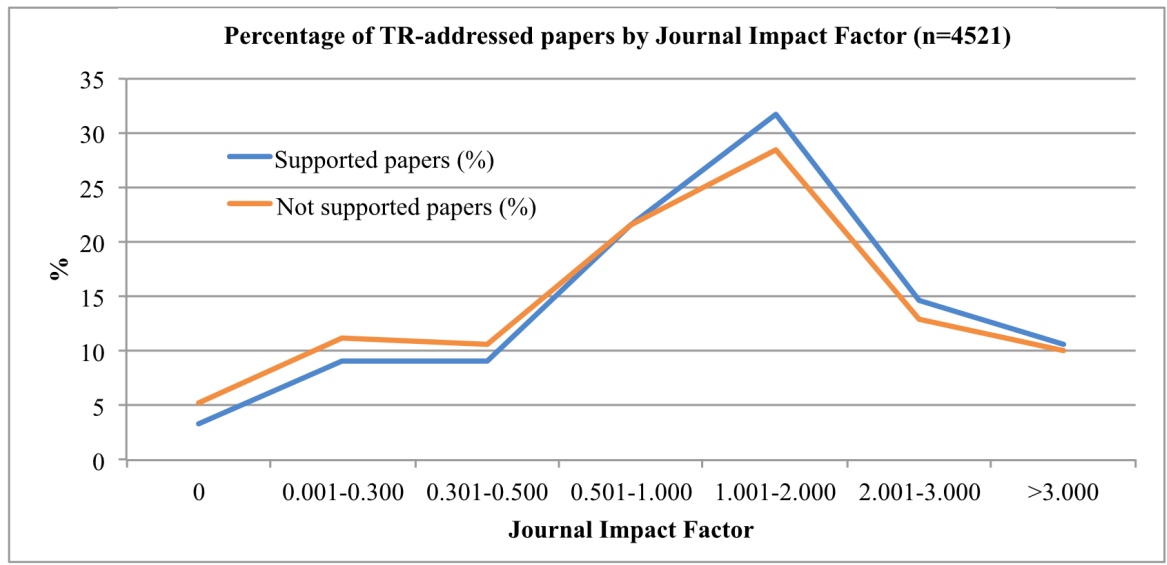

Fig. 4. Percentage of TR-addressed papers by Journal Impact Factor (n=4521)

We also looked at the distribution of the TR-addressed papers by JCR quartiles to find out if supported papers were published in journals listed in higher JCR quartiles under their respective subject categories. First, we analyze the distribution of the TR-addressed papers published in 2015 by JCR quartiles and compare it with that of JCR data representing all papers published in the same year.

Recall that over 40% of publications gets published in the high impact Q1 journals in comparison to that of about 15% in the low impact Q4 journals. However, the distribution of the TR-addressed papers published in 2015 by JCR quartiles is quite different: 21%, 20%, 23%, and 31% of them appeared in Q1, Q2, Q3, and Q4 journals, respectively (Table 3).[7] The percentage of TR-addressed papers published in high impact Q1 journals is less than half the percentage of papers published in all Q1 journals listed in JCR. In contrast, the percentage of TR-addressed papers published in low impact Q4 journals is twice that of all JCR Q4 journals.

The difference is even more striking when one considers the distribution of publications (articles and reviews) by quartiles according to 2016 JCR data (representing journals published in 2015). Some 44%, 27%, 16%, and 13% of all articles and reviews listed in Science Citation Index Expanded (SCI-E) in 2015 were published in Q1, Q2, Q3, and Q4 journals, respectively (Liu, Guo, & Zuo, 2018). The distribution by quartiles was somewhat similar for articles and reviews indexed under SCI-E's largest 25 research areas representing more than half of all publications in the database (Miranda & Garcia-Carpintero, 2019). The corresponding percentages for TR-addressed Science papers are as follows: 22% in Q1, 20% in Q2, 24% in

---

[6] Not all journals in which TR-addressed papers were published had both JIF and/or AIS values listed in JCR. The correlation coefficient is based on 3,961 papers with both values. Papers that were published in journals with no JIS and/or AIS were also excluded.

[7] Note that 4% of all TR-addressed papers were published in journals with no assigned JCR quartiles in 2015 (i.e., journals with no JIFs).

Q3, and 31% in Q4 journals (Table 3).[8] The percentage of TR-addressed papers published in low impact Q4 journals is more than thrice that of all JCR Q4 journals. Note that our data does not include reviews, although it is unlikely that the distribution of reviews by journal quartiles would differ much from that of articles in general, and that a small percentage (*ca.* 2%) of TR-addressed reviews would significantly change the distribution of journals by quartiles in this study in particular.

Table 3. Distribution of TR-addressed papers by JCR quartiles (n=4521)

| Subject | Papers | Quartiles | | | | | | | | | | Total* | |
|---|---|---|---|---|---|---|---|---|---|---|---|---|---|
| | | N/A | | Q1 | | Q2 | | Q3 | | Q4 | | | |
| | | N | % | N | % | N | % | N | % | N | % | N | % |
| **Science** | Supported | 26 | 2 | 391 | 26 | 332 | 22 | 329 | 22 | 430 | 29 | 1508 | 101 |
| | Not supported | 102 | 4 | 495 | 19 | 494 | 19 | 633 | 25 | 851 | 33 | 2575 | 100 |
| | *Subtotal* | *128* | *3* | *886* | *22* | *826* | *20* | *962* | *24* | *1281* | *31* | *4083* | *100* |
| **Social Sciences** | Supported | 29 | 17 | 24 | 14 | 38 | 22 | 35 | 21 | 45 | 26 | 171 | 100 |
| | Not supported | 44 | 17 | 41 | 15 | 48 | 18 | 50 | 19 | 84 | 32 | 267 | 101 |
| | *Subtotal* | *73* | *17* | *65* | *15* | *86* | *20* | *85* | *19* | *129* | *29* | *438* | *100* |
| **All** | Supported | 55 | 3 | 415 | 25 | 370 | 22 | 364 | 22 | 475 | 28 | 1679 | 100 |
| | Not supported | 146 | 5 | 536 | 19 | 542 | 19 | 683 | 24 | 935 | 33 | 2842 | 100 |
| | **Total** | **201** | **4** | **951** | **21** | **912** | **20** | **1047** | **23** | **1410** | **31** | **4521** | **99** |

\* Some total percentages in the rightmost column differ from 100% due to rounding.

We observe a somewhat similar relationship for TR-addressed papers listed in SSCI: 36%, 29%, 20%, and 15% of publications listed in SSCI were published in Q1 through Q4 journals, respectively) (Liu, Guo, & Zuo, 2018), whereas the corresponding percentages for TR-addressed papers are 15% in Q1, 20% in Q2, 19% in Q3, and 29% in Q4 journals. That the percentage of TR-addressed Social Science papers published in low impact Q4 journals is relatively lower (29%) compared to that of TR-addressed Science papers (39%) is due to the fact that an additional 17% of TR-addressed Social Science papers were published in journals with no JCR quartiles assigned.

Fig. 5 provides the JCR quartile distributions of all TR-addressed papers published in journals listed in JCR in 2015 (left panel) as well as that of supported and not supported ones (right panel). The percentage of papers decreases as we go from low impact Q4 journals to high impact Q1 journals for both Science and Social Science papers. Some 58% and 65% (including the percentage of papers published in journals with no quartiles assigned) of the TR-addressed Science and and Social Science papers were published in low impact Q3 and Q4 journals whereas only 42% and 35% of Science and Social Science papers, respectively, were published in high impact Q1 and Q2 journals (Fig. 5, left panel). The opposite is the case for all publications (articles and reviews) listed in SCI-E and SSCI: About 70% and 65% of all Science and Social Science papers, respectively, were published in high impact Q1 and Q2 journals (Liu, Guo, & Zuo, 2018; Miranda & Garcia-Carpintero, 2019).

As for the comparison of the distributions of supported and not supported TR-addressed papers by JCR quartiles, some 53% and 64% of supported Science and Social Science papers, respectively, were published in low impact Q3 and Q4 journals or in journals with no JIF values (hence no quartiles) (Fig. 5, right panel). The percentage of supported Social Science papers that appeared in journals with no (zero) JIF values (17%) was much higher than that of Science papers (2%) because Social Science papers got supported more generously to increase their number (Tonta, 2017b). Almost half (48%) the supported Science papers appeared in Q1 and Q2 journals (as opposed to 36% of supported Social Science papers). The percentage of supported Science papers published in Q1 journals (26%) is almost twice that of Social Science papers (14%). The percentage of supported Social Science papers with no quartiles was the same as those with no JIFs (17%).

---

[8] Note that 3% of all TR-addressed Science papers indexed in SCI were published in journals with no assigned JCR quartiles in 2015 (i.e., Science journals with no JIFs).

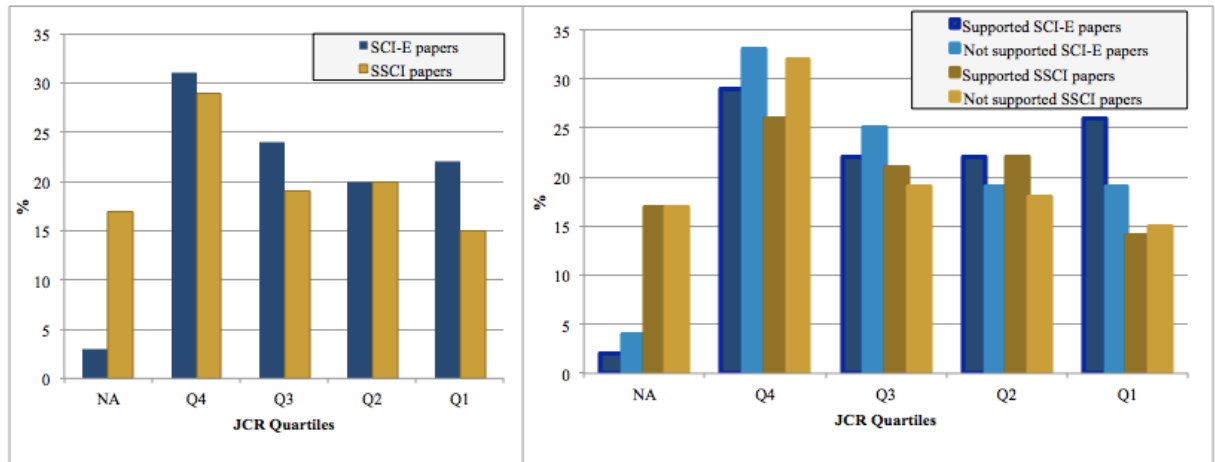

Fig. 5. JCR quartile distributions of TR-addressed papers. Left panel: all papers; right panel: supported and not supported papers (based on figures in Table 3 above).

The difference between quartile distributions of supported and not supported Science papers was statistically significant ($X^2(4) = 39.6$, $p < .05$), although the effect size was rather small ($r = .01$). This may be due to the more restrictive support policy towards Science papers published in the lowest quartile of journals (Tonta, 2017b, pp. 23-24). The percentages of supported papers by quartiles suggest that the support system seems to have failed to be more selective, thereby rewarding more papers that were published in journals with lower JCR quartiles and thus lower citation impact. This indicates that countries thinking of introducing graded incentives based on journal quartiles need to reconsider their plans.

The main findings of this study with regards to about 226,000 TR-addressed papers published between 2006 and 2015 are as follows: They were published in relatively low impact journals. More than half appeared in journals with AIS values well below the world average of AIS value for all journals (1.000) indexed in WoS. TR-addressed papers did not get cited very often, as half the papers received between zero (13%) and five citations within the study period. Supported and not supported papers collected comparable number of citations per paper. They were not significantly different from each other in terms of JIFs, AISs, and quartiles of journals in which they were published. The distributions of the supported and not supported Science and Social Science papers by citation impact values did not differ much, either.

Findings of the current study corroborate the findings of earlier studies of PRFSs to some extent (Auranen & Nieminen, 2010; Butler, 2003, 2004; Checchi, Malgarini, & Sarlo, 2019; Geuna & Martin, 2003; Good et al., 2015; Osuna, Cruz-Castro, & Sanz-Menéndez, 2011; Quan, Chen, & Shu, 2017). We have not observed a negative correlation between JIFs, AISs, and quartiles of supported and not supported TR-addressed papers. Yet, the average number of citations per paper and JIF, AIS and quartile values were quite similar for both supported and not supported papers, indicating that the support system of TÜBİTAK has not increased the citation impact of TR-addressed papers, which confirms the findings of an earlier study (Tonta, 2017b).

Some of the limitations of our study should also be noted. We lack empirical data as to why the support system did not have any considerable effect in increasing the citation impact of TR-addressed papers. One reason might be that researchers have had to write papers for tenure and academic promotion anyway, and they do not pay much attention to TÜBİTAK's support program (Yuret, 2017). Or, they may have found support through other sources (e.g., project budgets or other academic incentives). Yet another reason might be that the amount of support (based primarily on JIF values) was perhaps not attractive enough for some researchers, especially when we consider the fact that many TR-addressed papers were published in relatively low impact journals and the total amount per paper has to be divided by the number of co-authors (Tonta, 2017b).

As our findings are based on the citation impact of supported and not supported TR-addressed papers in general, we do not know if monetary support has actually incentivized some researchers to publish in high impact journals. For this, support data needs to be matched with the authors' data so that the research output of supported authors can be compared with that of not supported ones in terms of citation impact of their papers.

**Concluding Remarks**

So, does monetary support increase the citation impact of scholarly papers? This does not seem to be the case in our study. Findings indicate that both supported and not supported TR-addressed papers were somewhat similar in terms of average number of citations per paper. They have been published in journals with similar JIFs, AISs, and JCR quartiles. Contrary to the expectations of the research funders, payments transferred to researchers through the support program do not seem to have played much role in improving the citation impact of TR-addressed papers. This suggests that subsidies based on bibliometric measures function poorly as incentives to increase the quantity and quality of the scholarly papers.

The support system seems to have rewarded the authors of papers who published in mediocre or low impact journals relatively more often. Despite comparatively lower "piece rates" paid for papers published in such outlets, many researchers sought financial support nonetheless. This might be an indication that subsidies may have encouraged some researchers to develop "opportunistic behavior" and they may have acted like "rent seekers" interested only in reaping the relatively modest monetary benefits of the support program without much consideration for the quality of their papers, a conjecture begging further research.

In conclusion, the support program of TÜBİTAK and similar academic incentives of universities and the Turkish Higher Education Council should be reconsidered. Also, more comprehensive studies need to be carried out to find out why increasingly more researchers in Turkey are inclined to publish in low impact journals.

**Acknowledgements**

We thank Mr. M. Mirat Satoğlu of TÜBİTAK for providing data for supported papers, and Dr. Umut Al of Hacettepe University for reviewing an earlier version of this paper.